\documentclass[twocolumn]{revtex4}
\usepackage{amssymb}
\usepackage{latexsym}
\usepackage{epsfig}
\usepackage{color}                            

\usepackage{amsmath,amssymb}
\usepackage{graphicx} 
\usepackage{gastex}

\begin{document}
\title{New explanation for accelerated expansion and flat galactic rotation curves}
\author{Ahmad Sheykhi\footnote{asheykhi@shirazu.ac.ir}}
\address{Physics Department and Biruni Observatory, Shiraz University, Shiraz 71454, Iran\\
Institut f\"{u}r Physik, Universit\"{a}t Oldenburg, Postfach 2503
D-26111 Oldenburg, Germany}

 \begin{abstract}
Employing the non-additive Tsallis entropy, $S\sim A^{\beta}$, for
the large-scale gravitational systems, we disclose that in the
cosmological scales both Friedmann equation and the equation of
motion for the Newtonian cosmology get modified, accordingly. We
then derive the modified Newton's law of gravitation which is
valid on the large scales. We show that, in the relativistic
regime, the modified Friedmann equation admits an accelerated
expansion, for a universe filled with ordinary matter, without
invoking any kind of dark energy, provided the non-extensive
parameter is chosen $\beta<1/2$. In the non-relativistic regime,
however, the modified Newton's law of gravitation can explain the
flat galactic rotation curves without invoking particle dark
matter provided $\beta \lesssim1/2$. Our study may be regarded as
an alternative explanation for the ``dark side of the universe",
through modification of the gravitational field equations.

\end{abstract}
 \maketitle

 \newpage
\section{Introduction\label{Intro}}
It is quite possible to speculate that the observed astrophysical
and cosmological phenomenons such as the late time acceleration of
the universe expansion, the flat rotation curves of spiral
galaxies, the observed dynamics of the cluster of galaxies, the
gravitational lensing, which cannot be understood through the
standard Newton and Einstein theories of gravitation, are just
weakness of the underlying theory of gravity. If that is true, one
may expect that such  phenomenons are just the geometrical effects
arising due to the flaw of the theory. Therefore, one should be
capable to explain the observed cosmological phenomenons through
modifying the underlying theory of gravity. Many attempts have
been done to find possible solutions for the puzzles of
accelerated expansion and flat galactic rotation curves from
geometrical perspective. In particular, over the past years,
modified theories of gravity have gained considerable attentions.
Among them, perhaps the most known, is $f(R)$ gravity which try to
explain the early inflation, the late time acceleration and even
the flat galactic rotation curves through modification of
Einstein-Hilbert action (see
\cite{Cap,Noj,NOdi,NO1,Riazi,Sobuti,Cog,Sot,Chr,Sho,NO2,Cap2,NO3,Jian,Odi1,Odi2}
and references therein). For a recent comprehensive review on
different modified gravity techniques which contains all the
necessary information in the context of cosmology, emphasizing on
inflation, bouncing cosmology and late-time acceleration, we refer
to \cite{NO4}.

Another attempt for explanation of the dark matter puzzle, through
a non-relativistic model, is the modified Newtonian dynamics
(MOND) \cite{Milgrom}, which proposed to address the flat rotation
curves of spiral galaxies through modifying Newton's law of
gravitation. Although MOND theory can explain the flat galactic
rotation curves, however it suffers from several problems. First
of all, it is problematic to embed MOND theory within a more
comprehensive relativistic theory of gravity and hence, its
theoretical origin remains unclear. Second, it predicts that the
individual halo associated with a galaxy is infinite in extent,
while recent galaxy-galaxy lensing results suggest that galaxy
halos may have a maximum extent of about $0.5$ Mpc \cite{halo}.
Some authors have also tried to explain the flat rotation curves
through modification of Einstein gravity
\cite{Sobuti1,Sobuti2,Mim1,Myr2,MimMOND,ASJ}.

In this paper, we would like to propose and alternative
perspective for tackling the problem of accelerated expansion as
well as the flat galactic rotation curves through modifying
Friedmann equation and Newton's law of gravitation, using
thermodynamic argument. In $1902$ Gibbs pointed out that, in
systems where the partition function diverges, the standard
Boltzmann-Gibbs theory is not applicable, and large-scale
gravitational systems are known to fall within this class. Hence,
the usual Boltzmann-Gibbs additive entropy must be generalized to
the non-additive (non-extensive) entropy (the entropy of the whole
system is not necessarily the sum of the entropies of its
sub-systems) \cite{Tsallis,Lyra,Wilk}. In $1988$ Tsallis
generalized standard thermodynamics to non-extensive one, which
can be applied in all cases, and still possessing standard
Boltzmann-Gibbs theory as a limit \cite{Tsallis}. Based on this,
and using the statistical arguments, Tsallis and Cirto argued that
the entropy of a black hole does not obey the area law and can be
modified as \cite{Tsa}
\begin{equation}\label{TEnt}
S=\gamma A^{\beta},
\end{equation}
where $A$ is the horizon area, $\gamma$ is an unknown constant,
and $\beta$ known as nonextensive parameter.

On the other hand, thermodynamical interpretation of gravitational
field equations, based on the profound connection between the
first law of thermodynamics on the boundary, and the gravitational
field equations in the bulk, is now an established fact
\cite{Jac,Eling}. It was argued that, given an entropy expression
at hand, in any gravity theory, one can rewrite the Friedmann
equations of the Friedmann-Robertson-Walker (FRW) universe in the
form of the first law of thermodynamics on the apparent horizon
and vice versa
\cite{Wang,Fro,Pad1,PPad,CaiKim,Cai2,Cai3,Cai4,Cai5,Shey1,Shey2,Pad2,Shey3,JSS,JS}.
{Recently, when the entropy of the gravitational system is in the
form of non-extensive Tsallis entropy, the thermodynamical
interpretation of the gravitational field equation as well as the
effects of non-extensive parameter in the context of cosmology
have arisen a lot of interests. It was shown that non-extensive
parameter change the strength of the gravitational constant and
consequently the energy density of the dark components of the
universe, requiring more (less) dark energy to provide the
observed late time universe acceleration \cite{Barb,Nunes}. In the
context of the non-extensive Kaniadakis statistics \cite{Kan}, the
Jeans length was investigated and the results were compared with
the Jeans length obtained in the non-extensive Tsallis statistics
\cite{Abero}. The cosmological scenarios based on the
non-extensive Tsallis entropy have been explored in \cite{Em}. It
was shown that the Universe exhibits the usual thermal history,
with the sequence of matter and dark energy eras, and depending on
the value of non-extensive parameter, this scenario may exhibits a
varieties of dark energy models \cite{Em}. Taking the entropy
associated with the apparent horizon of the FRW universe in the
form of the Tsallis entropy, and assuming the first law of
thermodynamics, $dE=TdS+WdV$, holds on the apparent horizon, the
modified Friedmann equations describing the dynamics of the
universe with any spatial curvature were extracted \cite{SheT}. It
was argued that with appropriate choice for the non-extensive
parameter, this model is capable to reproduce the late-time cosmic
acceleration, as well as the early deceleration, in the absence of
dark energy \cite{SheT}. The studies on the Tsallis cosmology were
also generalized to the case with variable non-extensive parameter
\cite{NOS1}. It was shown that the extra terms, arisen from
non-extensive entropy, can play the role of an effective dark
energy describing the evolution of the universe from early epoch
to the late time acceleration \cite{NOS1}.  More recently, it was
shown that it is quite possible to establish a correspondence
between the non-extensive Tsallis cosmology and cosmology with a
fluid with redefined equation of state for both constant and
variable non-extensive parameter \cite{NOS2}. In this viewpoint,
the effective fluid can derive successfully not only the present
acceleration but also the early inflation without spoiling the
correct late-time acceleration \cite{NOS2}.}

In this paper, we consider the Tsallis entropy given in
(\ref{TEnt}) as the entropy expression of the gravitational
systems, and disclose that, in the relativistic regime, it could
modify Friedmann equation and the resulting equation is capable to
provide, naturally, the late time accelerated expansion without
invoking any kind of dark energy. On the non-relativistic regime,
however, the modification to the Newton's law of gravitation leads
to the explanation of the flat rotation curves of galaxies without
needing to particle dark matter. {Let us stress the similarity and
difference of the present work with alternative theories of
gravity, in particular $f(R)$ gravity. First of all, similar to
our work, $f(R)$-gravity theories can also explain the dark side
of the universe through modification of the geometrical part of
the Einstein gravity. For a comprehensive and excellent review on
$f(R)$-gravity we refer to \cite{NO2}, where it was shown that
$f(R)$ theory can be considered as a unified description for the
history of the universe from the early-time inflation to the
late-time acceleration. However, in $f(R)$-gravity, in order to
establish a correspondence between the field equations and the
first law of thermodynamics, a treatment with nonequilibrium
thermodynamics is required \cite{Eling,Cai5}. In this case the
first law of thermodynamics acquires an additional entropy
production term grown up internally due to the non-equilibrium
treatment of the system. This is in contrast to the Einstein
gravity, and also its modification when the entropy of the system
is modified as non-extensive Tsallis entropy. It was shown that,
in the presence of non-extensive entropy, the Friedmann equation
of FRW cosmology in a nonflat \cite{Em,SheT} and flat
\cite{NOS1,NOS2} universe can be deduced from the first law of
thermodynamics, on the apparent/Hubble horizon, in a complete
equilibrium situation. This is one of the main difference between
our work and $f(R)$ gravity. Besides, for explanation the flat
galactic rotation curves, we apply the non-relativistic modified
Newton's law of gravity based on the Tsallis entropy, while in
$f(R)$-gravity, the problem is addressed through modifying General
Relativity \cite{Sobuti,Chr,Sho}.}

This paper is outlined as follows. In the next section, we show
how the modified Friedmann equation in Tsallis cosmology leads to
the late time accelerated expansion in the Universe filled with
ordinary baryonic matter. In section III, we derive the modified
Newton's law of gravitation based on the Tsallis entropy. In
section IV, we show that the modified Newton's law of gravitation
can also be extracted from entropic force scenario. In section V,
we employ the modified Newton's law of gravity and disclose that
it can explain the flat rotation curves of spiral galaxies. In
this viewpoint, dark matter has only geometrical effect that
originates with a modification of gravity. The last section is
devoted to conclusions.

\section{Accelerated universe in Tsallis cosmology\label{FIRST}}
Let us start by deriving the modified Friedmann equation based on
the non-extensive Tsallis entropy from the first law of
thermodynamics. This problem was already studied in
\cite{Em,SheT,NOS1,NOS2}, but since it provides a basis for the
next sections, for completeness, we review the derivation here,
briefly. Following \cite{SheT}, we assume the non-extensive
Tsallis entropy affects on the geometry part of the Friedmann
equations and hence we keep the energy content of the universe in
the form of the standard perfect fluid. It is important to note
that a different viewpoint was, recently, adopted in \cite{NOS2}
by assuming that the non-extensive entropy modifies the energy
density and hence pressure of the universe. As a result, one
should redefine the equation of state of the perfect fluid
\cite{NOS2}.

Suppose the background of spacetime is given by the FRW geometry,
\begin{equation}
ds^2={h}_{\mu \nu}dx^{\mu}
dx^{\nu}+\tilde{r}^2(d\theta^2+\sin^2\theta d\phi^2).
\end{equation}
In the above line element $\tilde{r}=a(t)r$, $x^0=t, x^1=r$, and
$h_{\mu \nu}$=diag $(-1, a^2/(1-kr^2))$ stands for the metric of
two dimensional subspace. We also assume our universe is bounded
by an apparent horizon with radius $
 \tilde{r}_A={1}/{\sqrt{H^2+k/a^2}}.
$ Using the definition of surface gravity $\kappa$ for the
apparent horizon, it is easy to show that the temperature on the
apparent horizon can be given through relation \cite{Cai2}
\begin{equation}\label{T}
T=\frac{\kappa}{2\pi}=-\frac{1}{2 \pi \tilde
r_A}\left(1-\frac{\dot {\tilde r}_A}{2H\tilde r_A}\right).
\end{equation}
Suppose the energy-momentum tensor of the universe is $
T_{\mu\nu}=(\rho+p)u_{\mu}u_{\nu}+pg_{\mu\nu}, $ the conservation
equation, $\nabla_{\mu}T^{\mu\nu}=0$, for the FRW geometry implies
the continuity equation as $\dot{\rho}+3H(\rho+p)=0$. Because our
universe is expanding, as a thermodynamical system, a work should
be done due to the volume change of the system. The density of
this work, on the FRW background, is given by \cite{Hay2}
\begin{equation}\label{Work2}
W=\frac{1}{2}(\rho-p).
\end{equation}
Finally, we propose the first law of thermodynamics holds on the
apparent horizon,
\begin{equation}\label{FL}
dE = T dS + WdV.
\end{equation}
Note that for a pure de-Sitter space, $\rho=-p$, the the first law
reduces to $dE = T dS -pdV$. If we denote the total energy of the
universe $E=\rho V$ where $V=\frac{4\pi}{3}\tilde{r}_{A}^{3}$,
after differentiating, we arrive at
\begin{equation} \label{dE1}
 dE=4\pi\tilde
 {r}_{A}^{2}\rho d\tilde {r}_{A}+\frac{4\pi}{3}\tilde{r}_{A}^{3}\dot{\rho} dt.
\end{equation}
Substituting $\dot{\rho}$ from the continuity equation, yields
\begin{equation}
\label{dE2}
 dE=4\pi\tilde
 {r}_{A}^{2}\rho d\tilde {r}_{A}-4\pi H \tilde{r}_{A}^{3}(\rho+p) dt.
\end{equation}
Then, we should consider the evolution of the Tsallis entropy
which we assume has the form (\ref{TEnt}). Taking the differential
of the Tsallis entropy, we find
\begin{equation} \label{dS}
dS= 8\pi \gamma \beta (4 \pi {r}_{A}^2 )^{\beta-1} \tilde {r}_{A}
d\tilde {r}_{A}.
\end{equation}
Inserting Eqs. (\ref{T}), (\ref{Work2}), (\ref{dE2}) and
(\ref{dS}) in the first law (\ref{FL}), we obtain
\begin{equation} \label{Fried1}
\frac{\gamma \beta}{\pi \tilde {r}_{A}^3} \left(4\pi \tilde
{r}_{A}^2\right)^{\beta-1} d\tilde {r}_{A}= H(\rho+p) dt.
\end{equation}
Using the continuity equation, we get
\begin{equation} \label{Fried2}
-\frac{2}{\tilde {r}_{A}^3} \left(4\pi \tilde
{r}_{A}^2\right)^{\beta-1} d\tilde {r}_{A} = \frac{2\pi }{3\gamma
\beta}d\rho.
\end{equation}
Integrating yields
\begin{equation} \label{Frie3}
\frac{1}{\tilde {r}_{A}^{4-2\beta}}= \frac{2\pi (2-\beta)
}{3\gamma \beta} \left(4\pi \right)^{1-\beta}  \rho,
\end{equation}
where the constant of integration is set equal to zero. Finally,
we define the constant $\gamma$ as,
\begin{equation}\label{gamma}
\gamma\equiv\frac{2-\beta }{4\beta L_p^2 } \left(4\pi
\right)^{1-\beta},
\end{equation}
where $L_p=\sqrt{G\hbar/c^3}$ is the Planck length. Since the
entropy is positive definite ($\gamma>0$), the above definition
also implies $\beta<2$. After using definition $\tilde {r}_{A}$,
Eq. (\ref{Frie3}) immediately transforms to
\begin{equation} \label{Fried4}
\left(H^2+\frac{k}{a^2}\right)^{2-\beta} = \frac{8\pi L_p^2} {3}
\rho.\end{equation} In this way we derive the modified Friedmann
equation which describes the evolution of the universe in Tsallis
cosmology based on the non-extensive Tsallis entropy. When
$\beta=1$, the standard Friedmann equation is recovered. The
second Friedmann equation can be easily derived by combining the
continuity equation with Eq. (\ref{Fried4}).
Now we want to study the cosmological consequences of the obtained
modified Friedmann equation. It is a matter of calculation to show
that the second derivative of the scale factor satisfies the
following equation
\begin{eqnarray}
\frac{\ddot{a}}{a}
\left(H^2+\frac{k}{a^2}\right)^{1-\beta}=-\frac{4\pi
L_{p}^{2}}{3(2-\beta)} \left[(2\beta-1) \rho +3p\right].
\label{2Fri5}
\end{eqnarray}
Therefore, the accelerated expansion ($\ddot{a}>0$) can be
achieved provided,
\begin{eqnarray}
 (2\beta-1) \rho +3p <0, \  \    \Rightarrow  \  \  \omega< \frac{1-2\beta}{3},          \label{w1}
\end{eqnarray}
where $\omega=p/\rho$ denotes the equation of state parameter.
Condition (\ref{w1}) has interesting consequences. Let us consider
it carefully in two cases. In the first case, where
$\beta\geq1/2$, we always have $\omega< 0$ as a condition for an
accelerated universe. In the second case where $\beta<1/2$, it is
quite possible to have $\omega\geq0$, while our universe is still
accelerating ($\ddot a>0$). This is a very interesting result
which confirms that, in the framework of Tsallis cosmology, the
current acceleration of the universe expansion can be understood,
in the presence of the ordinary matter with $w\geq 0$. Precisely
speaking, we can consider a universe filled with pressureless
baryonic matter, and still enjoys an accelerated expansion without
invoking any dark companion for its matter/energy content.

The above discussion can also be confirmed by looking explicitly
to the scale factor. Assuming a flat FRW universe filled with
pressureless matter ($p=0$), the Friedmann equation (\ref{Fried4})
admits the solution $a(t)\sim t^{(4-2\beta)/3}$. This implies that
$ \ddot{a}(t)\propto (2-\beta)(1-2\beta) \ t^{-(2+2\beta)/3}, $
where the constant of proportionality is also a positive definite
\cite{SheT}. Thus, for $\beta<1/2$ we have an accelerated universe
($\ddot a>0$), in accordance with condition (\ref{w1}). It was
also argued, not only the accelerated expansion but also the early
deceleration as well as the age problem of the universe can be
circumvented automatically in the context of Tsallis cosmology
without invoking additional dark component of the energy
\cite{SheT}.

{We emphasize here that the authors of \cite{NOS1,NOS2} argued
that the modified Friedmann equations, derived from Tsallis
entropy, can reproduce the late-time acceleration provided one
take an effective dark-energy \cite{NOS1} or a redefined fluid
with generalized equation of state \cite{NOS2}. They also
established a correspondence between the modified cosmology
through non-extensive thermodynamics and the holographic dark
energy model as well as $f(R)$ gravity \cite{NOS1}. It is
important to note that for derivation the Friedmann equations, the
authors of \cite{NOS1,NOS2} assumed the first law of
thermodynamics on the Hubble horizon as $dQ=TdS$, where $dQ$ is
the heat flux crossing the horizon, and the spacetime is spatially
flat. While, here we derived the modified Friedmann equation for
any spatial curvature by assuming $dE=TdS+WdV$ holds on the
apparent horizon. Another difference between our work and
\cite{NOS1,NOS2} is that we could reproduce the late-time
acceleration in the presence of ordinary matter, without needing
to redefine the fluids or taking into account effective dark
energy.}

{Finally, it is worthy to note that the correspondence between
Tsallis cosmology and the standard Firedmann equation with the
fluids of redefined equation of stat established in \cite{NOS2},
comes from the fact that the field equations of General
Relativity, and hence the Friedmann equations, relate the geometry
of spacetime to its energy content. Thus, any modification of the
geometry can be translated to modification of the energy content
and vice versa. In Tsallis cosmology based on none-extensive
entropy, it is quite possible to consider the modification in the
geometry part of the gravitational field equation, and keep the
energy content as standard perfect fluid \cite{SheT}. This is
reasonable, because the definition of the entropy is based on the
area (geometry) of the system, and thus any modification in
entropy should affect the geometry part of the field equations and
vice versa \cite{Eling,Shey1,Shey2}.}
\section{Modified Newton's law of gravity\label{NL1}}
In this section we first derive the equation of motion describing
the evolution of the universe in Newtonian cosmology. Using this
equation, we then derive the Modified Newton's law of gravitation
which is based on the non-extensive Tsallis entropy. We start from
the Friedmann equation (\ref{Fried4}) by taking the time
derivative of it. We arrive at
\begin{eqnarray} \label{FrN1}
(2-\beta) \frac{\ddot{a}}{a}
\left(\dot{a}^2+k\right)^{1-\beta}=-\frac{4\pi L_{p}^{2}}{3}
\left[(2\beta-1) \rho +3
p\right]a^{2-2\beta}. \nonumber \\
\end{eqnarray}
When $\beta=1$, it reduces to
\begin{eqnarray} \label{FrNb1}
\frac{\ddot{a}}{a}=-\frac{4\pi L_{p}^{2}}{3} \left( \rho +3
p\right),
\end{eqnarray}
which is the evolutionary equation for the scale factor in
standard cosmology. We also assume in the Newtonian cosmology the
spacetime is Minkowskian with $k=0$, and work in the unit
$\hbar=c=1$, and so $L_{p}^{2}=G$ . Therefore, Eq. (\ref{FrN1})
reduces to

\begin{eqnarray} \label{FrN2}
(2-\beta) \frac{\ddot{a}}{a}=-\frac{4\pi G}{3} \left[(2\beta-1)
\rho +3p\right] \left(\frac{a}{\dot{a}}\right)^{2-2\beta}
\end{eqnarray}
We consider a compact spatial region $V$ with a compact boundary
$\mathcal S$, which is a sphere with physical radius $R= a(t)r
=H^{-1}$, where $r$ is a dimensionless co-moving coordinate which
remains constant for any cosmological object partaking in free
cosmic expansion. The active gravitational mass in General
Relativity, inside the volume $V$, is defined as \cite{Cai4}
\begin{equation}\label{ActM}
\mathcal M =2
\int_V{dV\left(T_{\mu\nu}-\frac{1}{2}Tg_{\mu\nu}\right)u^{\mu}u^{\nu}}.
\end{equation}
A simple calculation gives
\begin{equation}\label{ActM2}
\mathcal M =(\rho+3p)\frac{4\pi}{3}R^3.
\end{equation}
In order to transform from General Relativity to the Newtonian
gravity, we also replace the active gravitational mass $\mathcal
M$ with the total mass $M= \rho V=4\pi \rho R^3/3$. This is equal
to transforming $\rho+3p \rightarrow \rho$ in Eq. (\ref{FrN2}),
\begin{eqnarray} \label{FrN3}
(2-\beta) \frac{\ddot{a}}{a}=-\frac{4\pi G}{3}\rho(2\beta-1)
\left(\frac{a}{\dot{a}}\right)^{2-2\beta}.
\end{eqnarray}
This is nothing but the modified dynamical equation describing the
evolution of the universe in Newtonian cosmology. In the limiting
case where $\beta=1$, we find
\begin{eqnarray} \label{FrNb3}
\frac{\ddot{a}}{a}=-\frac{4\pi G}{3}\rho,
\end{eqnarray}
which is the standard equation of motion in Newtonian cosmology.
On the other hand, the acceleration of a test particle $m$ near
the surface $\mathcal S$ can be written
\begin{equation}\label{FN1}
\ddot{R}=\ddot{a} r=F/m.
\end{equation}
where $F$ is the gravitational force between $m$ from $M$.
Equating $\ddot{a}$ in Eqs. (\ref{FrN3}) and (\ref{FN1}), we find
\begin{eqnarray} \label{FrN4}
F=- \left(\frac{2\beta-1}{2-\beta}\right)\frac{4\pi G}{3}\rho m R
\left(\frac{a}{\dot{a}}\right)^{2-2\beta}
\end{eqnarray}
Using the fact that $R=1/H=a/\dot{a}$ and $\rho=M/V$, the above
equation can be rewritten as
\begin{eqnarray} \label{FrN5}
F=- \left(\frac{2\beta-1}{2-\beta}\right)\frac{G M m}{R^{2
\beta}}.
\end{eqnarray}
In this way we derive the modified Newton's law of gravity based
on the non-extensive Tsallis entropy. When $\beta=1$, one recovers
the well-known Newton's law of gravitation.
\section{Newton's law from entropic force\label{NLentropic}}
Now we want to employ the idea of entropic gravity proposed by
Verlinde \cite{Ver} and show that the Newton's law of gravity get
modified when the entropy of the system is in the form of Tsallis
entropy. According to the Verlinde's hypothesis, gravity can be
regarded as an entropic force caused by changes in the information
associated with the positions of material bodies. Using the first
principles, namely the equipartition law of energy in statistical
mechanics together with the holographic principle, he derived
Newton's law of gravitation, Poisson equation and in the
relativistic regime the Einstein field equations of General
Relativity \cite{Ver}. Although it was already addressed by
Padmanabhan \cite{Pad0} that gravity has a statistical origin, and
in particular, one can use the equipartition law of energy to
provide a thermodynamic interpretation of gravity, the notion that
gravity is not a fundamental force and can be identified as an
entropic force was first pointed out by Verlinde \cite{Ver}.

According to Verlinde, when a test particle moves apart from the
holographic screen, it will experience an effective force equal to
\begin{equation}\label{F}
F\triangle x=T \triangle S,
\end{equation}
where $T$ and $\triangle S$ are, respectively, the temperature and
the entropy of the surface and $\triangle x$ is the displacement
of the test particle from the holographic screen. Thus, in order
to have a non-vanishing entropic force, we need to have a non-zero
temperature.

Suppose a holographic screen, which by assumption is a spherically
symmetric surface $\mathcal {S}$ with area $A=4 \pi R^2$, is a
storage device for information and the holographic principle
holds. It is naturally to assume the total number of bits $N$, is
proportional to the area/entropy, $N \sim A \sim S$. The total
energy $E$ of the system inside the holographic screen is
distributed on these bits and thus the temperature on the surface
is given by the equipartition law of energy
\begin{equation}
\label{E}
 E=\frac{1}{2}Nk_B T  \  \Rightarrow \  T=\frac{2E}{N k_B}.
 \end{equation}
We also assume the total energy of the system can be written as
$E=Mc^2$ where $M$ is the total mass distribution inside the
holographic spherically symmetric screen which is uniformly
distributed \cite{Ver}. The surface $\mathcal {S}$ is located
between the test mass $m$ and the mass distribution $M$, and the
test mass is assumed to be very close to the surface compared to
its reduced Compton wavelength $\lambda_m={\hbar}/{(mc)}$.
Finally, we write the number of bits on the holographic surface as
\begin{equation}
\label{NA} N=4S=\frac{2-\beta}{\beta G} (4 \pi)^{1-\beta}
A^{\beta},
 \end{equation}
where we have assumed $S=\gamma A^{\beta}$ and used definition
(\ref{gamma}). Following \cite{Ver}, we postulate the change of
the entropy associated with the information on the holographic
screen equals
\begin{equation}
\label{deltaS} \triangle S =2 \pi k_B   \   \   \ \rm {when} \ \ \
|\triangle x|= \eta \lambda_m,
 \end{equation}
where $\eta$ is a constant which should be defined latter. We also
assume the entropy gradient points radially from the outside of
the surface to inside. Combining relations (\ref{E}), (\ref{NA})
and (\ref{deltaS}) with (\ref{F}) and working in the unit
$\hbar=c=1$, we arrive at
\begin{equation}
F= -\frac{\beta}{\eta (2-\beta)} \frac{GMm}{R^{2\beta}}.
 \end{equation}
Finally, we redefine $\eta=\beta/(2\beta-1)$ and rewrite the above
equation in the form
\begin{equation} \label{MNL}
F= -\frac{(2\beta-1)}{(2-\beta)} \frac{GMm}{R^{2\beta}}.
 \end{equation}
This is the modified Newton's law of gravitation derived from the
viewpoint that gravity is an entropic force, by assuming that the
entropy of a gravitational system is in the form of the
non-additive Tsallis entropy. It is clear that our result from the
entropic force approach coincides with the result obtained in the
previous section. Our investigation shows that with the correction
to the area law, the Newton's law of gravitation get modified,
accordingly.

It is important to note that in order to arrive at the standard
Newton's law of gravitation, one should take the limit
$\beta\rightarrow 1$, which corresponds to the area law for the
entropy. This implies that the modified Newton's law derived from
Tsallis entropy holds on the large-scales gravitational system,
namely at outside the galaxies. This is consistent with the
arguments which states that in systems with diverging partition
function, such as large-scale gravitational systems, the standard
Boltzmann-Gibbs theory cannot be applied and one needs to use
non-extensive, Tsallis thermodynamics, which still possesses
standard Boltzmann-Gibbs theory as a limit
\cite{Tsallis,Lyra,Wilk}. Clearly, $\beta\rightarrow 1$ is the
limiting case of our theory. Note that inside the galaxy, the
standard Newton's law can correctly predicate the orbital speed of
star and the problem of flat rotation curves (difference between
observation and theory) appears only for outside the galaxies. On
the cosmological scales, including outside the galaxies, the
entropy still obeys the non-extensive one, thus Tsallis cosmology
holds on the theory is realistic.

In the next section, we shall use this modified Newton's law of
gravity and show that it is indeed capable to explain the flat
rotation curves of spiral galaxies.
\section{Explanation of the galaxies rotation curves\label{NL1}}
There are a lot of observational data which confirm that the
rotational velocity curves of all spiral galaxies are proportional
to the distance from the center $v\propto r$, for inside the
galaxy, and rotation curves usually remain \textit{almost} flat
far from galactic centers, typically beyond $30-40$ kpc. For
inside the galaxy, however, these observations can be completely
understood via Newtonian gravity. Unfortunately, for outside a
spiral galaxy, Newton's law of gravitation is not capable to
explain the rotational curves, and there is indeed a contradiction
between observation and the prediction of the theory, since
Newton's law of gravitation predicts that objects that are far
from the galaxy center have lower velocities $v \propto
{r^{-1/2}}$, while observations imply that the velocity curves
flattened out to $v\simeq constant$. It is well established that
the baryonic matter of galaxies does not provide sufficient
gravitation to explain the observed dynamics of the systems. The
most widely adopted way to resolve these difficulties is the dark
matter hypothesis which suggests that the mass of galaxies
continues to grow even when there is no luminous component to
account for this increase. According to this hypothesis, all
visible galaxies are surrounded by massive nonluminous matters.
Besides the dark matter proposal, alternative theories of
gravitation have been also speculated and debated for
justification of the flat rotation curves. As we mentioned in the
introduction, Milgrom \cite{Milgrom} tried to explain the flat
rotation curves of galaxies, through modifying Newton's law of
gravity (MOND), and represents it as a geometrical effect.
However, the MOND theory suffers from the flaws as regards the
theoretical origin.
\begin{figure}[htp]
\begin{center}
\includegraphics[width=7cm]{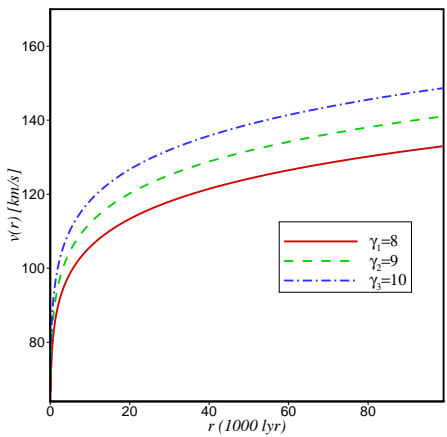}
\caption{The speed of a test particle around a spiral galaxy in
terms of distance for $\beta=0.40$ and different values of galaxy
mass $M_i=\gamma_{i} \times 10^9 M_{\odot}$.}\label{Fig1}
\end{center}
\end{figure}
\begin{figure}[htp]
\begin{center}
\includegraphics[width=7cm]{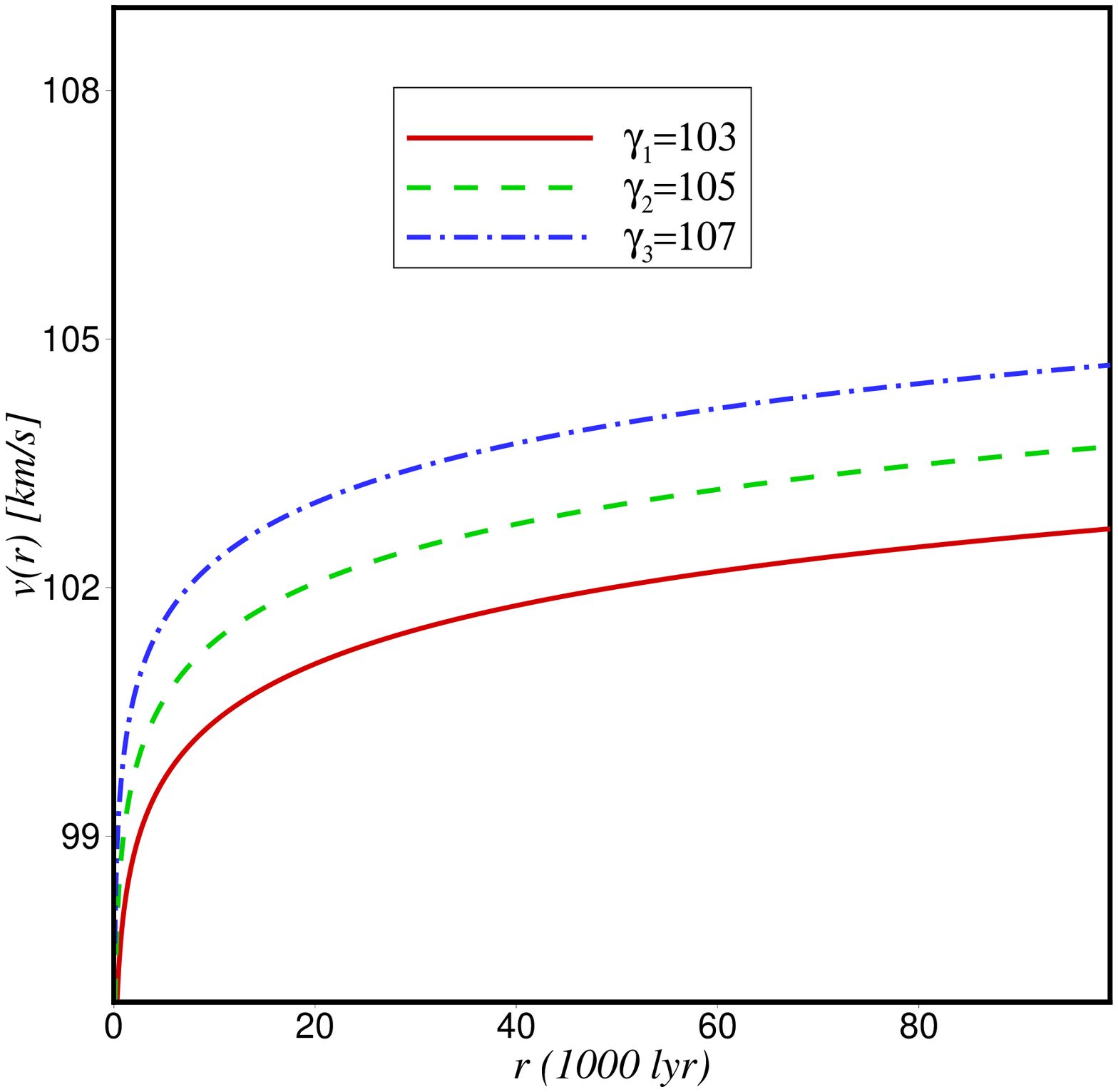}
\caption{The speed of a test particle around a spiral galaxy in
terms of distance for $\beta=0.49$ and different values of galaxy
mass $M_i=\gamma_{i} \times 10^9 M_{\odot}$.}\label{Fig2}
\end{center}
\end{figure}

\begin{figure}[htp]
\begin{center}
\includegraphics[width=7cm]{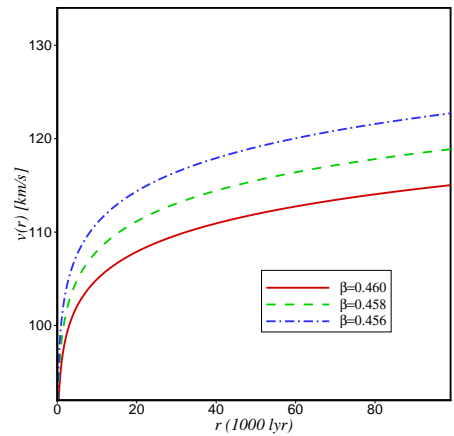}
\caption{The speed of a test particle around a spiral galaxy for a
typical galaxy with mass $M=25\times 10^9 M_{\odot}$.}\label{Fig3}
\end{center}
\end{figure}
\newpage
Here we tackle this problem by modifying Newton's law of gravity
based on the Tsallis entropy. Interestingly enough, we will show
that the modified Newton's law at large scales, given in Eq.
(\ref{MNL}), is capable to naturally explain the flat rotation
curves of spiral  galaxies. We postulate that $\beta=1$ for inside
the galaxies where the usual Newton's law holds, and $\beta<1/2$
at far distances, at galaxy out skirt. This implies that inside a
galaxy, and at distance $r$ from its center, we must have
\begin{equation}\label{MNL1}
a=\frac{v^2}{r}= \frac{GM(r)}{r^2} \rightarrow   \    v\propto r,
\end{equation}
where we have used the fact that inside the galaxy the total mass
which contributes to the velocity is $M(r)=4\pi r^3 \rho/3$.
However, at distances $r$ large enough for there to be no luminous
galactic component indicate that, we have $M(r)=M\simeq constant$,
and thus from Eq. (\ref{MNL}) we have
\begin{eqnarray}\label{MNL2}
&&\frac{v^2}{r}= \left(\frac{1-2\beta}{2-\beta}\right)
\frac{GM}{r^{2\beta}}  \nonumber \\
&& \Rightarrow v(r) =\sqrt{\left(\frac{1-2\beta}{2-\beta}\right)GM
r^{1-2\beta}}, \label{MNL3}
\end{eqnarray}
Clearly, $\beta\neq 1/2$, and in particular we should have $\beta
< 1/2$, but for latter convenience, we assume the values of
$\beta$ to be close to $1/2$. In order to have a better
understanding of the behaviour of $v$ in terms of distance, let
apply the above formula for a typical spiral galaxy. For this
purpose, we assume $M=M_i$ is the total mass of the galaxy. We
also set the Newtonian gravitational constant, $G \simeq
6.674\times 10^{-11} m^3 kg^{-1}s^{-2}$ and take the mass of the
Sun, $M_{\odot} \simeq 10^{30} kg $. Since the mass of a typical
galaxy is of order $\sim 10^9 M_{\odot}$, thus in these figures we
assume the mass of a typical galaxy ranges as, $8\times 10^9
M_{\odot} <M_i < 107 \times 10^9 M_{\odot} $, which are reasonable
values, at least for small spiral galaxies. We have plotted Figs.
1-3 for different values of the mass $M$ and non-extensive
parameter $\beta$. In all figures, we observe that the speed of a
test particle increase at small distance(inside galaxy) and tends
to almost a constant value at far distance, at galaxy out skirt.
From Figs. 1 and 2 we see that for a fixed value of $\beta<1/2$,
but close to it, at any distance, the orbital speed increases with
increasing the mass of the galaxy. Also, Fig. 3 one can see that
for fixed value of $M$, at any distance, the orbital speed
increases with increasing the non-extensive parameter $\beta$.
These figures are compatible with astrophysical data
\cite{NP,Man,Bri}. Note that here we have presented the ideas, and
have shown how the modified Newton's law of gravity given in
(\ref{MNL3}), at large distance, can explain the flat galactic
rotation curves. We leave the details of data fitting of this
model with observations for future studies.

\section{conclusions}\label{Con}
Using the non-extensive Tsallis entropy for the large-scale
gravitational systems, we disclosed that on the relativistic
cosmological background, the Firedmann equations describing the
evolution of the FRW universe get modified, accordingly. Starting
from the first law of thermodynamics on the apparent horizon, we
derived the modified Friedmann equation. We observed that when the
non-extensive parameter satisfies $\beta<1/2$, the late-time
acceleration of the cosmic expansion can be achieved in the
presence of the ordinary matter. This implies that one may
consider a universe filled with baryonic matter, and still enjoys
an accelerated expansion without invoking any dark companion for
its matter/energy content. On the other hand, in the regime of
non-relativistic gravity, one is able to reproduce the equation of
motion describing the evolution of the universe in Newtonian
cosmology. We then derived the modified Newton's law of
gravitation based on the Tsallis entropy from two approaches. The
first one is an inverse approach by starting from Newtonian
cosmology, and the second one is extracted through entropic force
scenario \cite{Ver}. We showed that both approaches lead to the
same results. Interestingly enough, we observed that flat galactic
rotation curves can be explained, through modified Newton's law of
gravitation provided $\beta \lesssim 1/2$, without needing to
particle dark matter.

Finally, we would like to stress that in this work, in contrast to
$f(R)$ gravity theories which try to explain the flat galactic
rotation curves through modifying Einstein gravity, we implemented
the problem in the context of non-relativistic modified Newton's
law of gravity. In a sense our work can be located in the MOND
theories category, however, the advantage of our work is that its
theoretical origin is well established.

\acknowledgments{I am grateful to anonymous referee for valuable
comments which helped me improve the paper significantly. I thank
Shiraz University Research Council. My special thanks go to Jutta
Kunz for helpful discussions and the University of Oldenburg, for
hospitality. }


\end{document}